\documentclass[twocolumn,runningheads]{svjour2}
\smartqed  
\usepackage{graphicx}
\journalname{Astrophysics and Space Science}
\begin{document}
\title{Gamma rays from halos around stars and the Sun
}
\author{E. Orlando         \and
        A. W. Strong }
\institute{E. Orlando \at
              MPE,Postfach 1312, D-85741 Garching, Germany \\
              \email{elena.orlando@mpe.mpg.de}           
           \and
           A. W. Strong \at
              MPE, Postfach 1312, D-85741 Garching, Germany \\
              \email{aws@mpe.mpg.de}  }
\date{Received: date / Accepted: date}
\maketitle
\begin{abstract}
Inverse Compton (IC) scattering by relativistic electrons produces  a major component of the diffuse  emission from the Galaxy.  The photon fields involved are the cosmic microwave background and the interstellar radiation field (ISRF)  from stars and dust. We expect  which could be detectable by instruments such as GLAST. Even individual nearby luminous stars could be detectable assuming just the normal cosmic-ray electron spectrum.
We present the basic formalism required and give possible candidate stars to be detected and make prediction for GLAST. Then we apply the formalism to the OB associations and the Sun, showing that the IC emission produced is not negligible compared to the sensitivity of current or coming detectors. 
We estimate that the gamma-ray flux from the halo around the Sun contributes to the diffuse background emission at few percent level.
\end{abstract}
\section{Introduction}
\label{intro}
In the early 90s there was already the idea to consider the IC gamma-ray emission generated by electrons accelerated by shocks in winds around hot stars (e.g.\cite{Ref0}). 
In the present work, for the first time, we show that even the gamma-ray emission by the ambient cosmic-ray electrons via IC scattering of the stellar radiation field could be detected by GLAST.
We begin with the simplest possible rough estimate to show that the IC emission from luminous stars could be visible.
The optical luminosity of the Galaxy is about 3$\times$10$^{10}$ L$_\odot$, and a typical O star has   10$^{5}$ L$_\odot$ i.e. about   10$^{-5}$ of the Galaxy.
Consider such a star at a distance of 100 pc: compared to the entire Galaxy (distance to centre = 8.5 kpc) this inverse Compton source is on average about a factor 100 closer and hence the IC is 10$^{-5}$ $\times$ 100$^2$ of the Galactic IC, suggesting it is significant. 
The IC luminosity $L_{IC}$ within a volume surrounding a star is proportional to the radius $r$ around the star times the optical luminosity of the star:
$L_{IC}\propto r~L_{STAR}$.
The flux depends on the star's distance d:
$flux_{IC}\propto L_{IC}/d^{2}$
and for an angle $\alpha \propto ~r/d$,
we obtain
$flux_{IC}\propto L_{STAR} ~\alpha/d$.
A more precise formulation is given in the next section.
\section{IC   single stars: theory}
\label{sec:2}
The gamma ray intensity as function of gamma energy of the scattered photon is:
\begin{equation}\label{eq:1}
I(E_{\gamma})=\frac{1}{4\pi}\int\epsilon(E_{\gamma})~ds
\end{equation}
where the emissivity $\epsilon$ is given by:
{\setlength\arraycolsep{2pt}
\begin{eqnarray}
\epsilon(E_{\gamma})=\int dE_{e}\int\sigma(\gamma,E_{ph},E_{\gamma})n_{ph}(E_{ph})cN(E_{e})dE_{ph}
\end{eqnarray}
where $N(E_{e})=AE_{e}^{-p}$ ~is the electron spectrum and $\sigma$ is the Klein-Nishina cross section for isotropic scattering \footnote{We have also computed the anisotropic case but the difference is less than 10$\%$.}.  $E_{ph}$ is the stellar photon energy and  $n_{ph}$ the photon density.
We assumed the star to be a black body where the energy density is characterized by the effective temperature  following the Stefan-Boltzmann equation. 
\begin{figure}[!h]
\centering
	\includegraphics[width=.35\textwidth, angle=270]{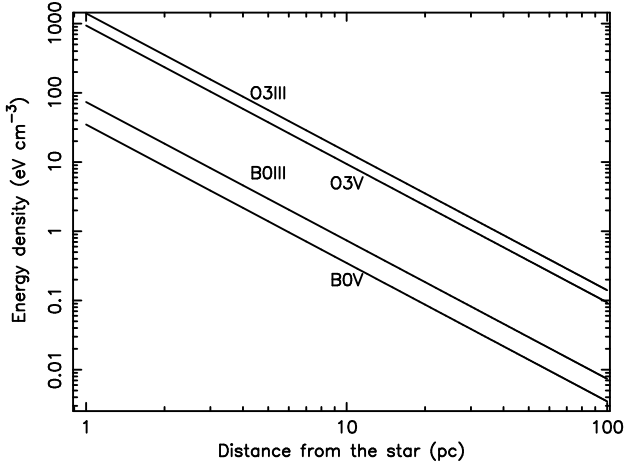}
	\caption{Photon energy density around different main sequence and giant stars from 1 pc to 100 pc.}
\label{orlando_fig1}
\end{figure}
The photon density at distance $r\gg R$ from the star is:
$n_{ph}(r)=(1/4)~n_{ph}(R)R^{2}/r^{2}$
since the energy ~density ~at~ $r$~ is ~ $(\sigma/c) T_{eff}^{4}(R^{2}/r^{2})=(a/4)~T_{eff}^{4}~(R^{2}/r^{2})$
where $R$ is the radius of the star, $n_{ph}(R)$ is the black body density at the effective temperature and $a$ is the radiation constant.  We assumed this relation for all distances. The value of the photon energy density  obtained for bright stars (Fig \ref{orlando_fig1}) is above the mean interstellar value of about 1~eV~cm$^{-3}$ even at 10 pc distance from the star, suggesting it contributes to clumpiness in the ISRF and hence in the IC emission. 
In order to obtain the inverse Compton radiation over a line-of-sight at an angle $\alpha$ from the star, the density variation over the line has to be known. 
\begin{figure}[!h]
\centering
	\includegraphics[width=8.3cm, angle=0]{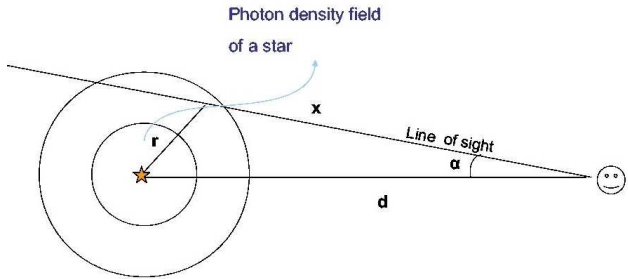}
	\caption{ Definition of variables for eq (3).}
\label{orlando_fig2}
\end{figure}
As presented in Fig \ref{orlando_fig2}, for a given point $x$ on the line-of-sight, the surface photon density is proportional to $1/r^{2}$, where $r^{2}=x^{2}+d^{2}-2~d~x~cos\alpha$, with $d$ distance from the star. Integrating the photon density over the line of sight from $x=0$ to $x=\infty$ 
and over the solid angle, the total photon flux produced by inverse Compton becomes:  
{\setlength\arraycolsep{2pt}
\begin{eqnarray}
f(E_{\gamma})=\frac{1}{4\pi} \int_{0}^{2\pi} d\varphi \int_{0}^{\alpha} sin\alpha ~d \alpha \int dE_{ph} {}
\nonumber\\
{}\times \int \sigma_{KN}(\gamma,E_{ph}, E_{\gamma})~c~N(E_{e})~dE_{e}~ \frac{n_{ph}(R)}{4} ~R^{2}  {}
\nonumber\\
{}\times \int_{0}^{\infty}\frac{dx}{x^{2}+d^{2}-2~x~d~cos\alpha}= {}
\nonumber\\
{}=\frac{R^{2}}{16d}(\pi \alpha + (\frac{\pi}{2})^{2}-arctan^{2}(cot\alpha)){}
\nonumber\\
{} \times \int dE_{ph} \int \sigma_{KN}~c~N(E_{e})~n_{ph}(R)~dE_{e} {}
\end{eqnarray}
which for small $\alpha$ is proportional to $\alpha/d$ and the intensity $I$ (per solid angle) is proportional to 1/($\alpha d$).
\section{IC for stellar types and distances}
\label{sec:3}
The IC emission for stars has been computed using the electron spectrum shown in Fig \ref{orlando_fig3}. This approximates the local interstellar electron spectrum  based on direct measurements. Explaining the diffuse Galactic emission including the `GeV excess' requires about a factor 4 higher electron spectrum described by the optimized model \cite{Ref1}. The present estimate is therefore rather conservative.  
\begin{figure}[!h]
\centering
\includegraphics[width=0.35\textwidth, angle=270]{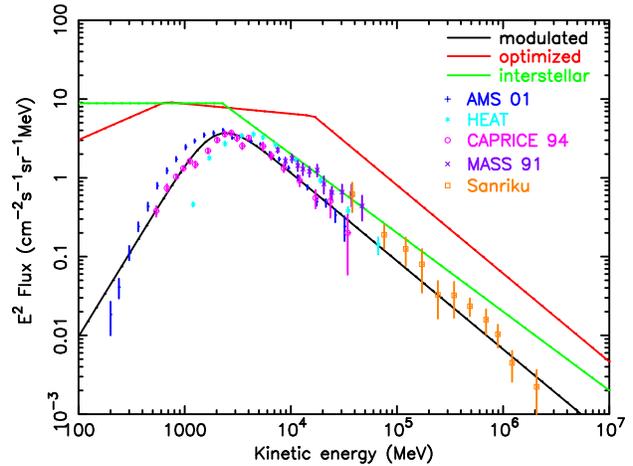}
\caption{Adopted modulated electron spectrum (black line), adopted interstellar model (green line) and optimized model (red line) compared with the data. Data: AMS 01 \cite{Ref2}, HEAT 94-95 \cite{Ref3}, CAPRICE 94 \cite{Ref4}, MASS 91 \cite{Ref5} and Sanriku \cite{Ref6}}
\label{orlando_fig3}
\end{figure}
The gamma-ray flux depends on the star distance and on the angle of integration: in fact it scales as the linear inverse of the distance and to first approximation is proportional to the angular distance from the star. Fig \ref{orlando_fig4} shows, as an example, the spectrum of the gamma-ray emission of stars of different spectral types and Fig \ref{orlando_fig5} the gamma-ray flux as function of the angle of integration \footnote{Note that the main contribution to the emission comes from more than about a pc from the star and hence is mostly beyond the influence of stellar winds \cite{Ref7}.}. 
\begin{figure}[!h]
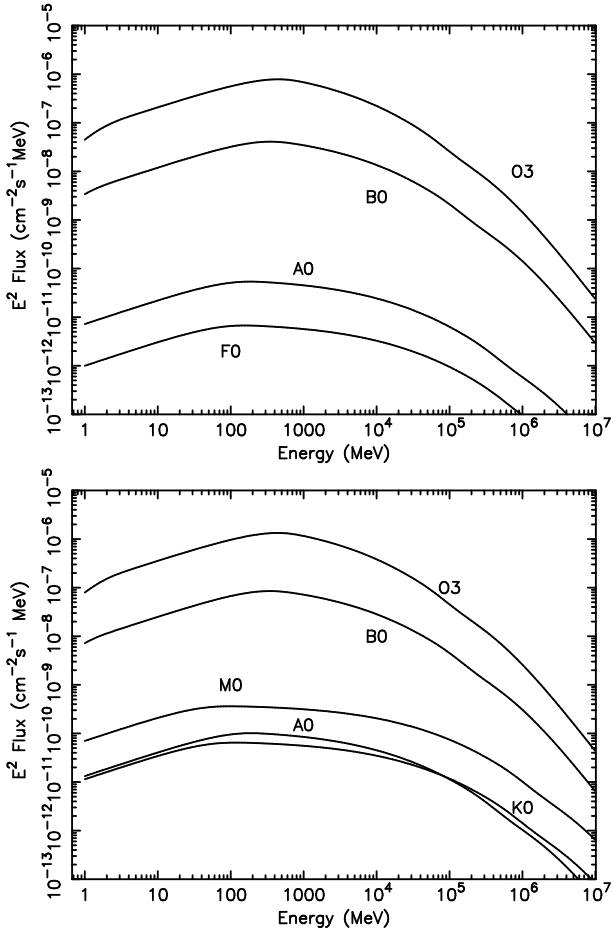

\centering
\includegraphics[width=0.35\textwidth, angle=270]{orlando_fig4a.ps}
\includegraphics[width=0.35\textwidth, angle=270]{orlando_fig4b.ps}
\caption{ Gamma-ray spectrum generated by IC scattering on the photon field of main sequence stars (upper) and giant stars (lower) at 100 pc distance. Flux is integrated over $5^{\circ}$ radius.}.
\label{orlando_fig4}
\end{figure}
\begin{figure}[!h]
\centering
\includegraphics[width=0.35\textwidth, angle=270]{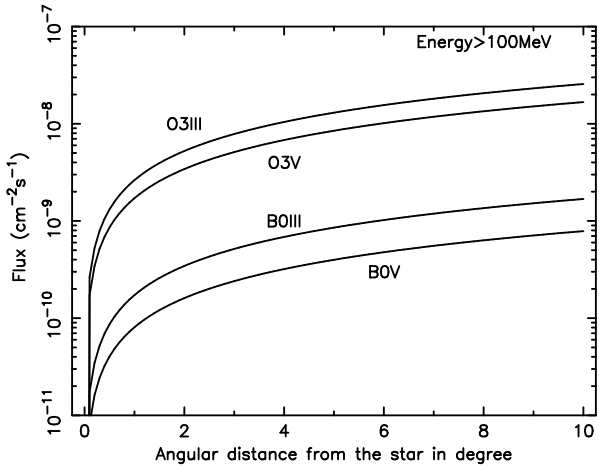}
\caption{Cumulative flux integrated over solid angle from stars at 100 pc distance as a function of angle for E$_{\gamma}$ $>$ 100 MeV.   }
\label{orlando_fig5} 
\end{figure}
\section{Candidates for detection and predictions for GLAST}
\label{sec:4}
The most luminous and nearby stars are candidates for giving a significant flux contributing to the total Galactic emission. Since the IC sources are extended, the angular radius to which the flux is taken is arbitrary, but we choose a value of 5$^{\circ}$ which is a compromise between angular resolution
and sensitivity of gamma-ray telescopes. The value of this angle could be further optimized for specific cases.
\subsection{Possible stellar candidates }
\label{sec:5}
In order to simulate the gamma-ray flux from halos around stars, the possible candidates have been taken from the Hipparcos nearby star catalogue \cite{Ref8}, choosing only the first 70 most luminous stars. This list includes stars up to 600 pc distance. The results are compared in Fig \ref{orlando_fig6} with the GLAST point source sensitivity \cite{Ref9} above 100 MeV of about 3$\times$10$^{-9}$cm$^{-2}$s$^{-1}$ for one year observation and about 1.5$\times$10$^{-9}$cm$^{-2}$s$^{-1}$ for 5 years observation. 
\begin{figure}[!h]
\centering
\includegraphics[width=0.35\textwidth, angle=270]{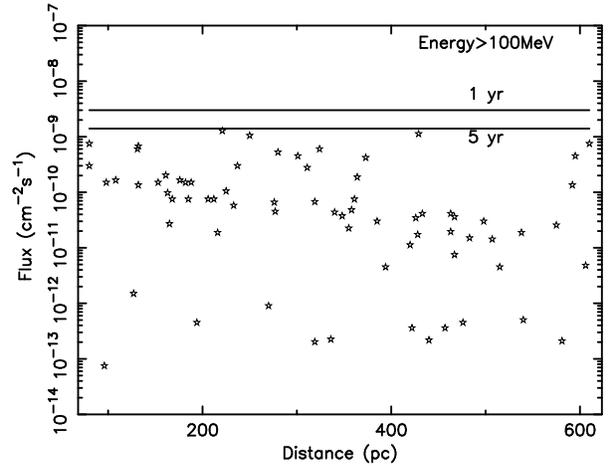}
\caption{Gamma flux integrated over 5$^{\circ}$ angle from the 70 most luminous stars from Hipparcos catalogue for E$_{\gamma}$ $>$ 100 MeV compared with the GLAST point source sensitivity for 1 year and 5 years observation (horizontal lines) }
\label{orlando_fig6}
\end{figure}
According to our prediction, the most gamma-ray bright stars are $\kappa$ Ori (B0Iab, distance 221 pc), $\zeta$ Pup (O5Ia, distance 429 pc, parameters from  \cite{Ref10}) and $\zeta$ Ori (O9.5Ib, distance 250 pc) with a flux respectively of 1.2, 1.1 and 1 $\times$ 10$^{-9}$cm$^{-2}$s$^{-1}$ for energy$>$ 100 MeV and within 5$^{\circ}$ angle. Further stars with a flux above 10$^{-10}$ cm$^{-2}$s$^{-1}$ are Betelgeuse, $\delta$ Ori, $\beta$ Ori or Rigel, $\zeta$ Per, $\lambda$ Ori, $\epsilon$ cMa. Another important candidate is $\eta$ Carinae with T$_{E}$ = 30000 K, 2.3 kpc distance and luminosity of 7$\times$ $10^{6}$ solar luminosity \cite{Ref11}. The estimated IC flux within 5$^{\circ}$ angle is 2.2, 0.1, 0.005 $\times$10$^{-9}$ cm$^{-2}$s$^{-1}$ respectively for energy $>$ 100 MeV, $>$ 1 GeV and $>$ 10 GeV.  However the fluxes obtained are affected by errors due to the big uncertainty of the parameters of the stars which could mean that some can be brighter $\gamma$-ray sources than these estimates.
\subsection{OB associations: Cygnus OB2 }
\label{sec:5}
Apart from individual stars, the full stellar population will exhibit features due to their clustering e.g. in OB associations. As an example the flux from Cygnus OB2 has been estimated for 120 O stars and 2489 B stars at 1700 pc distance. Data for the Cygnus association and stellar componets are taken from  \cite{Ref12}. A conservative assumption includes only O9 and B9 main sequence stars giving a flux within 5$^{\circ}$ of 4.8, 0.5, 0.02 $\times$ 10$^{-9}$ cm$^{-2}$s$^{-1}$ respectively for energy $>$ 100 MeV, $>$ 1 GeV and $>$~10~GeV, while a more realistic assumption includes all O6 and B5 main sequence stars, giving a flux of 18, 1.9, 0.05 $\times$10$^{-9}$ cm$^{-2}$s$^{-1}$ respectively. For 1$^{\circ}$ instead the estimated flux is 3.7, 0.3 and 0.008 $\times$10$^{-9}$ cm$^{-2}$s$^{-1}$ for the previous energy ranges and the more realistic assumption. This will clearly be of interest for GLAST. Furthemore we note that cosmic rays may also be accelerated in colliding winds in OB associations (eg. \cite{Ref13} and \cite{Ref14}) which would further increase the fluxes.
\section{The Sun}
\label{sec:6}
We apply the basic formalism to the Sun and find that the IC emission is not negligible. The IC emission has been computed using the modulated solar electron spectrum shown in Fig \ref{orlando_fig3}. Fig \ref{orlando_fig7} shows the IC intensity as a function of angular distance from the Sun compared with the EGB for different energy ranges. Fig \ref{orlando_fig8} shows the spectrum at different angles. We estimate that the gamma-ray intensity from the halo around the Sun contributes to the diffuse background emission a few percent of the extragalactic background (EGB) even at large angular distance from the Sun.  
\begin{figure}[!h]
\centering
\includegraphics[width=0.35\textwidth, angle=270]{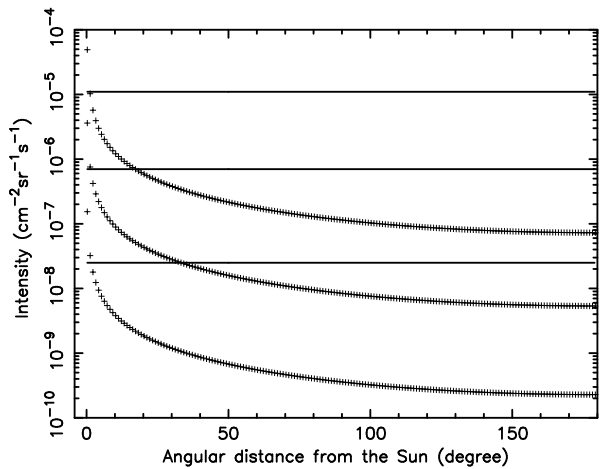}
\caption{ Estimated IC intensity (dotted lines) as a function of angular distance from the Sun compared with the EGB \cite{Ref15} (solid lines) for different energy ranges: top to bottom $>$ 100 MeV, $>$ 1 GeV and $>$ 10 GeV. }
\label{orlando_fig7}
\end{figure}
\begin{figure}[!h]
\centering
\includegraphics[width=0.35\textwidth, angle=270]{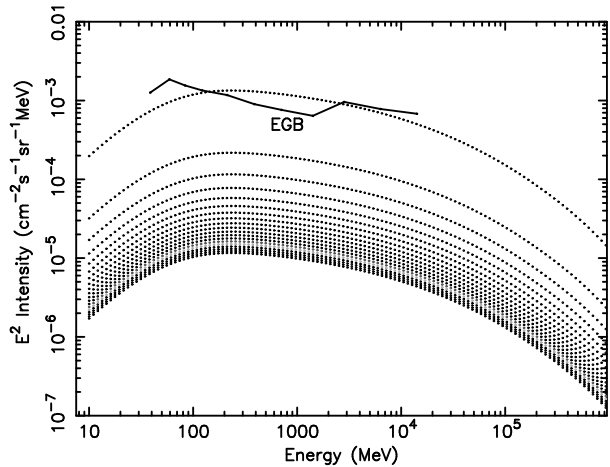}
\caption{ Gamma-ray spectrum for different angles from the Sun. Dotted lines show the spectra from top to bottom at steps of 5$^{\circ}$ from 1$^{\circ}$ to 90$^{\circ}$, while the solid line represents the EGB measured by EGRET \cite{Ref15}.}
\label{orlando_fig8}
\end{figure}
The estimated IC flux within 5$^{\circ}$ angle is 100, 8.4, 0.3 $\times$10$^{-9}$ cm$^{-2}$s$^{-1}$ respectively for energy $>$ 100 MeV, $>$ 1 GeV and $>$ 10 GeV, while within 1$^{\circ}$ angle it is 18, 1.3, 0.05 $\times$10$^{-9}$ cm$^{-2}$s$^{-1}$ for the same energy ranges. The gamma flux in the range 100 MeV $\div$ 10 GeV for 5$^{\circ}$ angle is about 5$\%$ Crab. 
The electron spectrum is however uncertain (Fig~\ref{orlando_fig3}) so that the IC flux could be larger or smaller by a factor $\sim$ 2.
In future a model of the gamma-ray flux from the Sun will be implemented, in order to take it into account for diffuse background emission studies. 
\section{Conclusion}
\label{sec:7}
We have estimated the gamma-ray emission by IC scattering of cosmic-ray electrons with the radiation field around stars. We find that the contribution of the most luminous stars is non-negligible and even individual luminous stars could be detectable by GLAST. Moreover OB associations can contribute to the clumpiness of the emission. The same model applied to the Sun \footnote{When this work had already been completed  we learned about work by Moskalenko et al. (2006) on the Sun \cite{Ref16}.} shows that the IC emission produced is significant and should be accounted for in diffuse background studies. 

\end{document}